\documentclass[final,12pt]{clear2023} 

\title[Learning Causal Representations of Single Cells via Sparse Mechanism Shift Modeling]{Learning Causal Representations of Single Cells \\via Sparse Mechanism Shift Modeling}
\usepackage{times}

\usepackage[utf8]{inputenc} 
\usepackage[T1]{fontenc} 
\usepackage{url}  
\usepackage{booktabs} 
\usepackage{amsfonts} 
\usepackage{nicefrac} 
\usepackage{microtype}  
\usepackage{xcolor} 
\usepackage{tikz}

\usetikzlibrary{bayesnet}
\usetikzlibrary{arrows}
\usepackage{color}
\usepackage{graphicx}
\usetikzlibrary{backgrounds}
\usepackage{float}
\usepackage{bm}
\usepackage{wrapfig}

\usepackage{mathtools}
\DeclarePairedDelimiterX{\infdivx}[2]{(}{)}{%
  #1\;\delimsize\|\;#2%
}
\newcommand{\infdiv}{D_{KL}\infdivx}

\usepackage{amssymb}
\usepackage{amsmath,amscd}
\usepackage{color}

\newcommand\blfootnote[1]{%
  \begingroup
  \renewcommand\thefootnote{}\footnote{#1}%
  \addtocounter{footnote}{-1}%
  \endgroup
}

\clearauthor{
\Name{Romain Lopez} $^\dagger$\blfootnote{$^\dagger$ These authors contributed equally to this work.}  \Email{lopez.romain@gene.com} \\
\addr Genentech Research and Early Development\\
\addr Stanford University \\
\AND
\Name{Nata\v{s}a Tagasovska}$^\dagger$ \Email{natasa.tagasovska@roche.com} \\
\addr Genentech Research and Early Development\\
\AND
\Name{Stephen Ra} \Email{ra.stephen@gene.com}\\
\addr Genentech Research and Early Development\\
\AND
\Name{Kyunghyun Cho} \Email{cho.kyunghyun@gene.com}\\
\addr Genentech Research and Early Development\\ 
\addr New York University\\
\addr CIFAR Fellow \\
\AND
\Name{Jonathan K. Pritchard} \Email{pritch@stanford.edu} \\
\addr Stanford University\\
\AND
\Name{Aviv Regev} \Email{regeva@gene.com}\\ 
\addr Genentech Research and Early Development\\
}

\begin{document}

\maketitle

\begin{abstract}
Latent variable models such as the Variational Auto-Encoder (VAE) have become a go-to tool for analyzing biological data, especially in the field of single-cell genomics. One remaining challenge is the interpretability of latent variables as biological processes that define a cell's identity. Outside of biological applications, this problem is commonly referred to as learning disentangled representations. Although several disentanglement-promoting variants of the VAE were introduced, and applied to single-cell genomics data, this task has been shown to be infeasible from independent and identically distributed measurements, without additional structure.
Instead, recent methods propose to leverage non-stationary data, as well as the sparse mechanism shift assumption in order to learn disentangled representations with a causal semantic. Here, we extend the application of these methodological advances to the analysis of single-cell genomics data with genetic or chemical perturbations. More precisely, we propose a deep generative model of single-cell gene expression data for which each perturbation is treated as a stochastic intervention targeting an unknown, but sparse, subset of latent variables. We benchmark these methods on simulated single-cell data to evaluate their performance at latent units recovery, causal target identification and out-of-domain generalization. Finally, we apply those approaches to two real-world large-scale gene perturbation data sets and find that models that exploit the sparse mechanism shift hypothesis surpass contemporary methods on a transfer learning task. We implement our new model and benchmarks using the scvi-tools library, and release it as open-source software at \url{https://github.com/Genentech/sVAE}.
\end{abstract}


\section{Introduction}
\label{sec:intro}
Machine learning methods have been key to gaining insights from large, high-dimensional genomic datasets, especially in single-cell genomics~\citep{ching2018opportunities}. Variational Auto-Encoders (VAEs)~\citep{kingma2013auto,rezende2014stochastic}, a recent approach in inferring complex data generative processes, are often well-suited for these applications because they allow for flexible model design, while keeping necessary changes to the inference procedure relatively minimal. Many generative models have been proposed for analyzing diverse biological data modalities, including gene (RNA) expression, chromatin accessibility and quantitative protein measurements~\citep{yau2019bayesian,lopez2020enhancing}.

However, VAEs suffer from a critical disadvantage due to their lack of interpretability, reflected as the absence of direct correspondence between individual latent variables and biological processes~\citep{way2018extracting}. While disentanglement-promoting VAEs~\citep{higgins2016beta,chen2018isolating+} can help better relate the two in genomics~\citep{eraslan2022single}, these methods often compromise the quality of the latent variables for downstream tasks~\citep{kimmel2020disentangling}. This is perhaps not surprising given recent theoretical developments showing that the recovery of ground truth latent variables is impossible from independent and identically distributed (i.i.d.) measurements~\citep{locatello2019challenging}. In the remainder of this paper, we interchangeably use the terms ``disentanglement'' and ``latent variable recovery'', following recent perspectives on non-linear Independent Component Analysis (ICA)~\citep{locatello2019challenging}.

Recent efforts in disentanglement instead focus on the assumption of non-stationary data~\citep{khemakhem2020variational}, where data must be (i) observed in different regimes, with known pairing between data points and regimes, (ii) generated such that regimes are incurring changes in latent variables, and (iii) latent variables are conditionally independent given the regime. In this configuration, latent recovery with a conditional VAE~\citep{sohn2015learning} is indeed theoretically possible. Follow-up work~\citep{lachapelle2022disentanglement} also draws connections to causal representation learning~\citep{scholkopf2022causality}, in which each regime may be modeled as an intervention targeting an unknown subset of latent variables. 

Recent advances in biotechnology made non-stationary data increasingly available, especially in the context of genetic or chemical perturbation screens with single-cell transcriptomic profiling as a readout~\citep{ji2021machine,peidli2022scperturb}. The Compositional Perturbation Autoencoder (CPA)~\citep{lotfollahi2021compositional} was introduced to embed perturbation profiles in latent space of an auto-encoder and predict the effect of unseen combinations of single perturbations. However, this method neither exploits the non-stationary assumption in its probabilistic model nor refers to any identifiability guarantees for the latent space. Thus, to the best of our knowledge, there have been no applications of the principles of disentanglement exposed in~\citet{khemakhem2020variational} and \citet{lachapelle2022disentanglement} to these new biological data types. 

\begin{figure}[t]
    \centering
    \includegraphics[width=\textwidth]{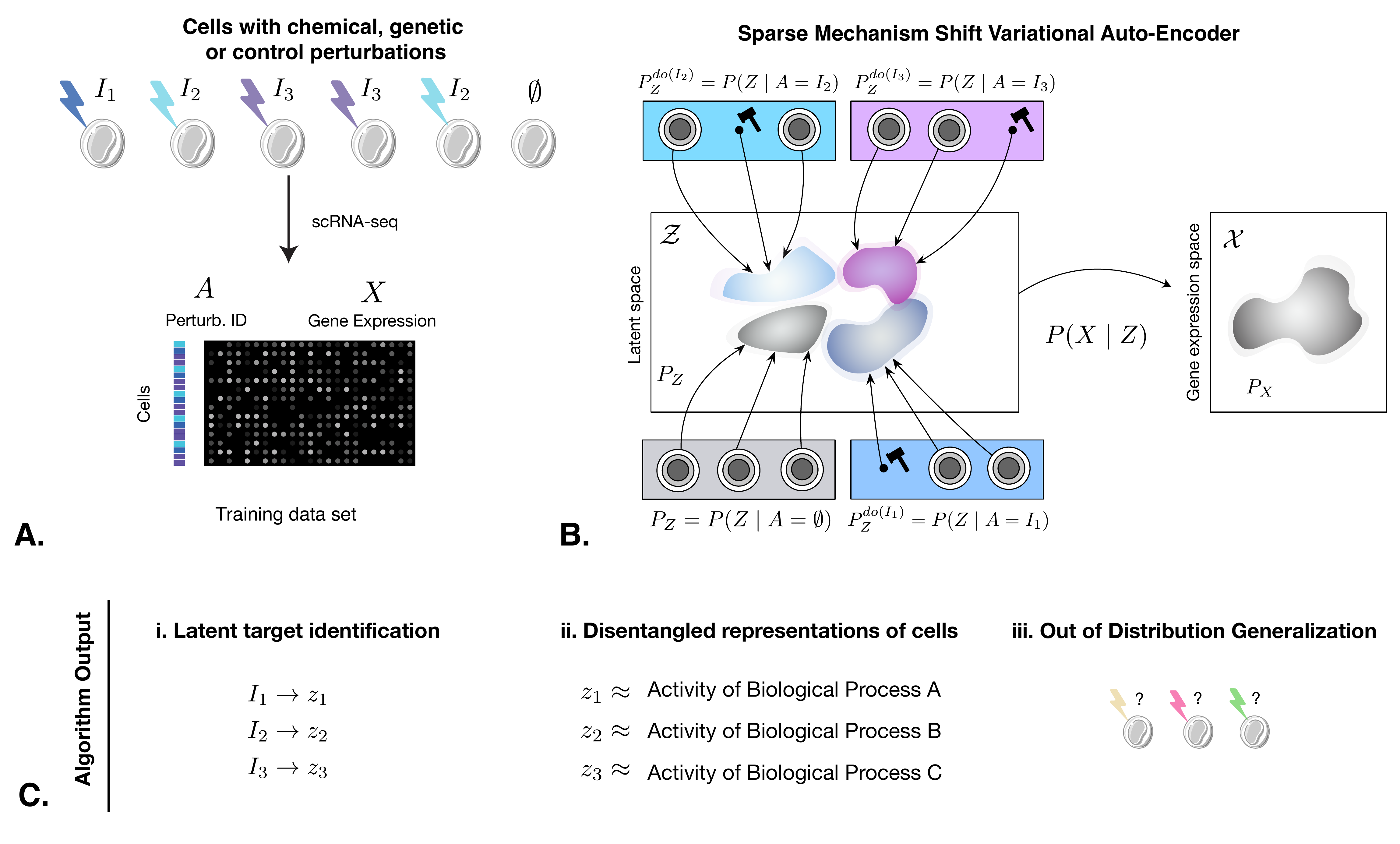}
    \caption{Overview of the sparse VAE framework applied to single-cell perturbation data. (A) Input data are gene expression profiles of cells under different genetic or chemical perturbations (colors), as well as the intervention label. (B) A schematic of the generative model, and the causal semantics of the sparse VAE (C) Three method outputs. (i) identification of target latent variables, encoded as a causal graph between the interventions and latent variables; (ii) a disentangled latent model for which individual latent variables are more likely to be interpreted as the activity of a relevant biological process; and (iii) the generalization of the generative model to unseen interventions (e.g., for latent target identification).}
    \label{fig:overview}
\end{figure}

Here, we explore real-world applications of those principles (Figure~\ref{fig:overview}). The promise is that causal models may eventually yield representations of perturbations and cells that are more mechanistically interpretable and more efficient for out-of-domain generalization. For example, the learned causal representations may lead to the delineation of biological processes such as gene programs, that are affected by perturbations~\citep{dixit2016perturb}. Furthermore, the model may also be used to project samples from unseen perturbations onto an existing atlas using transfer learning~\citep{lotfollahi2022mapping}.

After a brief introduction to disentangled representation learning and its intersection with causal inference (Section~\ref{sec:background}), we describe our motivation and effort to explore the learning of causal representations of perturbed cells profiled by single-cell RNA-seq (scRNA-seq) (Section~\ref{sec:ssm_sc}). We also introduce sVAE+, a variant of the sVAE~\citep{lachapelle2022disentanglement} with a Bayesian approach for learning sparse mechanism shifts that requires minimal hyperparameter tuning. Next, we introduce a benchmarking tool for simulations of single-cell perturbation data and the evaluation of algorithms for latent units recovery, intervention target recovery, and transfer learning (Section~\ref{sec:benchmark}). Finally, we present an application of the methods to two large-scale genetic screening experiments (Section~\ref{sec:real_data}). We show that models that exploit the sparse mechanism shift assumption outperform all methods by a significant margin on a transfer learning task. Our results suggest that causal inference is a promising paradigm for modeling the effects of perturbation in modern data sets from molecular biology.
\section{Background}
\label{sec:background}

This paper is concerned with the recovery of latent variables that initially generated the data, as well as providing a causal semantic to those latent variables. Therefore, we briefly introduce non-linear Independent Component Analysis (ICA), one of the prominent methods for accomplishing this task, and its relationship to causal representation learning. 

\subsection{Non-linear Independent Component Analysis}

ICA assumes that $x \in \mathbb{R}^d$ is generated using $p$ independent latent variables $z = (z_1, \ldots, z_p)$, called \emph{independent components} \citep{hyvarinen2002independent}. More precisely, observations $x$ are generated as $x = f(z) + \epsilon$ with $f$ a mixing function and $\epsilon$ an exogenous noise variable. The ICA literature focuses on the identifiable case (\textit{e.g.} if $f$ is a linear function, then the original $z$ may be recovered). In the general case of a non-linear \emph{mixing function} $f$ however, the model is unidentifiable from i.i.d. observations of $x$~\citep{hyvarinen1999nonlinear}.
Given this negative result, several papers introduced identifiable forms of non-linear ICA models~\citep{harmeling2003kernel, sprekeler2014extension, hyvarinen2016unsupervised, hyvarinen2017nonlinear}, based on the assumption that components $(z^i)_{i=1}^p$ are conditionally independent given an \emph{additional auxiliary} random variable $a \in \mathbb{R}^K$. 
Examples of auxiliary variables $a$ include the past components in the case of time series analysis or some form of class label~\citep{hyvarinen2016unsupervised}. With the observation of auxiliary variables, latent recovery is possible up to a linear transformation under sufficient conditions~\citep{hyvarinen2016unsupervised}.


The iVAE framework~\citep{khemakhem2020variational} proposes a VAE-based approach~\citep{kingma2013auto, rezende2014stochastic} for learning the parameters $\theta$ of a generative model $p_\theta(x \mid z)p_\theta(z \mid a)$, as well as $\phi$, those of a variational approximation $q_\phi(z \mid x, a)$ to the posterior $p_\theta(z \mid x, a)$. The iVAE specifies $p_\theta(z \mid a) = \mathcal{N}(\mu_a, I)$ as a Gaussian location-scale family with isotropic variance. As for the VAE, the parameters $(\theta, \phi)$ of the iVAE are learned via maximization of the evidence lower bound (ELBO): 
\begin{align}
  \log p_\theta(x \mid a) \geq 
  \mathbb{E}_{q_\phi(z \mid x, a)}\log \frac{p_\theta(x, z \mid a)}{q_\phi(z \mid x, a)}.
\end{align}
\label{eq:iVAE_elbo}

\subsection{Causal Inference from Unknown Interventions in Latent Space}

Recent theoretical work~\citep{lachapelle2022disentanglement, lachapelle2022partial} explores the assumption of sparse connections between the auxiliary variables $(a^l)_{l=1}^K$ and the latent components $(z^i)_{i=1}^p$, encoded in the form of a bipartite graph $G^a = ([K], [p], E)$, where $E$ is the edge set. In the case where $a$ describes a discrete data regime via one-hot encoding, the sparsity pattern of $G^a$ corresponds to the one of the mean vectors $\mu_a$ of $p_\theta(z \mid a)$. 

The novel sparsity assumption allows for recovery of latent units up to a permutation under weaker assumptions than~\citet{khemakhem2020variational}. Perhaps as importantly, it also allows for the interpretation of the graph $G^a$ from a causal perspective. More precisely, Theorem 22 of~\citet{lachapelle2022disentanglement} applies in the case where $a \in \lbrace e_1, e_2, \ldots, e_K \rbrace$, where each of $e_l$ for $l \in [K]$ is a one-hot vector encoding the $l$-th intervention, and each intervention has unknown targets on the set of components of $z$. The (unknown) graph $G^a$ describes which latent components are targeted by the intervention, that is $G_{i, l}^a = 1$ if and only if the $l$-th intervention targets $z^i$. In this context, the sparsity assumption corresponds precisely to the \emph{sparse mechanism shift} hypothesis from \citet{scholkopf2022causality} i.e. that only a few mechanisms change at a time.

The VAE variant introduced in~\citet{lachapelle2022disentanglement} (sparse VAE; sVAE) has an identical generative model as the one from the iVAE, except for the prior $p_\theta(z \mid a)$ for which a (stochastic) binary mask $\hat{G}^a_i \sim \textrm{Bernoulli}(\pi_i^a)$ is applied to the location parameter via element-wise product. The estimation procedure also relies on variational inference, with an addition of the regularization term $\alpha \| \pi_i^a\|_1$ to the ELBO, where $\alpha$ is a hyper parameter. To allow for gradient-based optimization of the objective function, sVAE uses the Gumbel-sigmoid distribution, a continuous relaxation of the Bernoulli distribution~\citep{jang2016categorical, maddison2016concrete}.


\section{A Sparse Mechanism Shift Model for Single Cell Measurements}
\label{sec:ssm_sc}

For self-containment and context of this work, we first describe why interventions are a sensible model for single-cell perturbations, and then propose a new model for causal representation learning of those data.

\subsection{Single-cell Perturbation Profiles as Interventional Data}

Experimental advances in biology now allow us to actively intervene and change the properties of a single cell by some genetic~\citep{norman2019exploring} or chemical~\citep{srivatsan2020massively} perturbation and then simultaneously profile each individually-perturbed cell for the identification of the perturbation and its molecular profile. A \emph{genetic perturbation} may be induced by the delivery of a guide RNA in a cell expressing a CRISPR-associated protein (e.g., Cas9). Upon delivery, the CRISPR complex performs a genetic intervention (e.g., knock-out) at the location of the gene targeted by the RNA guide, altering the function of the gene, along with the associated protein. In the case of a \emph{chemical perturbation}, cells are growth in the presence of a small molecule, which may enter the cell via the plasma membrane and interfere with one or several biochemical reactions. In both cases, interventions can affect (directly and indirectly) the expression of many genes (gene programs) that correspond to the activity of interpretable biological processes.

In such experiments, gene expression is profiled in each cell separately, after a fixed time, with single-cell RNA sequencing (scRNA-seq) --- a well-established technology used in diverse research areas of biology such as development~\citep{semrau2017dynamics}, autoimmunity~\citep{gaublomme2015single}, and cancer~\citep{patel2014single}. The measurements of a scRNA-seq experiment are summarized into a \emph{gene expression matrix} $X=(X_{1},\ldots,X_{N})\in \mathbb{N}^{N\times d}$, with $N$ the number of instances (cells), and $d$ the number of genes. Individual entries of this matrix $x_{ng}$ count the number of transcripts aligned to gene $g$ in cell $n$. The same scRNA-seq assay also captures the identity of each intervention in each cells, often in the form of an RNA-expressed barcode sequenced alongside native gene expression levels. This information is summarized into an \emph{intervention design matrix} $A\in\{0, 1\}^{N\times K}$, where $K$ denotes the total number of treatments (Figure~\ref{fig:overview}A).

\subsection{Generative Model under Sparse Mechanism Shift}

\paragraph{Perturbation Model} We assume a fixed number $p$ of latent variables $z = [z_1, \ldots, z_p]$, each representing the activity of a distinct biological process. Each intervention $a \in [K]$ targets latent variable $i \in [p]$ with probability
\begin{align*}
    \pi^a_{i} \sim \mathrm{Beta}\left(1, K\right),
\end{align*}
such that the binary variable
\begin{align*}
    \gamma^a_{i} \sim \mathrm{Bernoulli}\left(\pi^a_{i}\right),
\end{align*}
encodes whether latent variable $i$ is targeted by the intervention. Then, latent variable $z_i$ under intervention $a$ is generated as a mixture distribution
\begin{align}
\label{eq:prior}
    z_i \mid a \sim \gamma^a_{i} \mathrm{Normal}\left(\mu^a_{i}, 1\right) + (1-\gamma^a_{i}) \mathrm{Normal}\left(0, 1\right).
\end{align}
We note that the choice of the hyperpriors for the Beta distribution is sparsity inducing~\citep{moran2022identifiable}. Therefore, this generative model assumes that each perturbation should only affect a small subset of the latent variables. This hypothesis, inspired by~\citet{lachapelle2022disentanglement}, encourages the model to represent cells by the activity of biological processes that describe the effect of the perturbations. 

\paragraph{Measurement Model} For each single cell $n \in [N]$, we measure the gene expression vector $x_n = [x_{n1},\ldots, x_{ng}]$, as well as the perturbation information $a_n \in [K]$. Latent variables $z_n$ are generated conditionally on $a_n$, following~\eqref{eq:prior}. The expected frequency of expression of gene $g$ is calculated as
\begin{align*}
    \rho_{ng} = f(z_n),
\end{align*}
where $f$ is a neural network with two hidden layers, 128 hidden units at each layer, and ReLU non-linearity in between hidden layers. Notably, $f$ also has a softmax non-linearity at its output, to allow for interpretation of its output as a frequency of expression. Finally, gene expression $x_{ng}$ is generated as 
\begin{align*}
    x_{ng} \sim \mathrm{NegativeBinomial}\left(l_n \rho_{ng}, \theta_g\right),
\end{align*}
where $l_n$ is the number of RNA transcript captured in cell $n$ (referred to as library size, $l_n = \sum_{g}x_{ng}$), and $\theta_g$ are inverse-dispersion parameters. Indeed, the number of RNA transcripts captured in a single-cell is (mostly) treated as an artifact of the assay, and must be factored out of the learned representation. The choice of the negative binomial distribution is motivated by the fact that the data takes the form of counts, with overdispersion~\citep{grun2014validation}. All in all, this measurement model (a simplified version from the one described in~\citet{lopez2018deep})  takes into account major technical factors of variation in the data, and encourages the latent variables to learn patterns more reflective of the biological signal.

\paragraph{Connection to prior work} Existing non-linear ICA models such as the iVAE and the sVAE are not directly applicable to single-cell data, due to their unsuitable noise models. However, if we use the measurement model presented here, and change the prior on $\gamma^a_{i}$ to be a point mass at 1, we obtain a model assimilated to the iVAE. Similarly, if we place a Laplace prior on $\pi^a_i$ and perform MAP inference on $\gamma^a_i$, we obtain a model close to the sVAE. The model outlined here can be seen as a Bayesian treatment of the mechanism sparsity model in sVAE, and we therefore refer to it as sVAE+ in the remainder of this paper.

\subsection{Variational Inference}

The marginal probability of the data $p(x \mid a)$ is intractable. Therefore, we proceed to posterior approximation with variational inference to learn the model's parameters. We approximate the posterior distribution of each $\{\pi^a_{i}, \gamma^a_{i}, z_{ni}\}_{a \in [K], i \in [p], n \in [N]}$ with the mean-field variational distribution:
\begin{align}
    \bar{q} = \prod_{n \in [N]}q(z_n \mid x_n, a_n) \prod_{a \in [K], i \in [p]}q(\gamma^a_{i})q(\pi^a_{i}).
\end{align}
As in VAEs, each $q(z_n \mid x_n, a_n)$ follows a Gaussian distribution with diagonal covariance matrix. The parameters of those distributions are encoded via neural networks. Each of the latent variables $q(\gamma^a_{i})$ follows a Bernoulli distribution with free variational parameters. Finally, we use a point mass for each of $q(\pi^a_{i}) = \delta_{\psi^a_{i}}$, therefore performing MAP inference over this set of latent variables. We optimize the ELBO, derived as:
\begin{align}
    \label{eq:svae_obj}
\mathbb{E}_{\bar{q}}\left[\sum_{n=1}^N \log \frac{p(x_n, z_n \mid \gamma_{a_n})}{q(z_n \mid x_n, a_n)} + \sum_{a \in [K], i \in [p]}\log \frac{p(\gamma^a_{i}, \pi^a_{i})}{q(\gamma^a_{i})q(\pi^a_{i})}\right].
\end{align}

This objective function is amenable to stochastic optimization, as in~\citet{kingma2013auto}. This allows us to sample a fixed number of data points at each iteration, as well as from the variational distribution using the reparameterization trick and the Gumbel-sigmoid distribution for $q(\gamma^a_{i})$. We provide the derivation of~\eqref{eq:svae_obj} function in Appendix~\ref{app:elbo}. Additionally, we discuss practical challenges we encountered for training this model, alongside with implementation details in Appendix~\ref{app:implementation}. We implemented sVAE+ and the other baselines within the \verb |scvi-tools| codebase~\citep{gayoso2022python}. 

\subsection{Downstream utilization of sVAE+}


The sVAE+ model provides three main benefits (Figure~\ref{fig:overview}C). First, the estimated graph $G^a$ identifies which latent variables are affected by which perturbations. Edges in this graph are calculated by binarizing the matrix $(\pi^a_{i})_{a \in [K], i \in [p]}$ at threshold $0.5$. This helps in biological interpretation. For example, we can discover sets of perturbations that affect the same biological process. Second, the sparsity constraint on the latent space promises to help in disentanglement: identifying individual latent variables as distinct gene programs coordinately regulated by the perturbation, consistent with our understanding of the underlying organization of a cellular molecular circuits~\citep{heimberg2016low}. Third, an important byproduct of the causal semantic is that one can reasonably expect the learned representations to perform better at downstream tasks such as transfer learning~\citep{scholkopf2022causality}, a growing use-case of deep generative models in single-cell genomics~\citep{lotfollahi2022mapping}. In this work, we present a synthetic case of holding out perturbations and performing target identification with a fixed generative model. However, we anticipate that transfer learning will be helpful in other concrete tasks, such that the projection of new cells onto perturbation atlases~\citep{peidli2022scperturb} or the prediction of the outcome of perturbations across cellular contexts (e.g., different cell lines or tissues).

To the best of our knowledge, sVAE+ is the first proposed framework to explicitly model cellular perturbations as interventions on latent variables for the purpose of understanding causal mechanisms, while incorporating a model of experimental noise from single cell RNA-seq assays.

\section{A Sandbox for Evaluation of Learned Representations}
\label{sec:benchmark}


With sVAE+ being at the intersection of representation learning and single-cell biology, we consider it is important to present its alignment/compliance with both fields. To do so, we propose a sandbox for evaluation of current and future methods aimed at learning causal representations of single-cell data.  
Our sandbox consists of three main components: (i) simulation module for perturbed scRNA-seq data (ii) implementation of relevant baselines adapted for single-cell expression data and (iii) evaluation module including quantitative metrics for evaluation of learned representations.

\subsection{Sandbox overview} 

\paragraph{Simulated data} An important component of our sandbox is the simulation framework, that allows for systematic evaluation of the methods for all tasks on scRNA-seq data. We defer the details of the simulation framework in Appendix~\ref{app:simulation}, and briefly present here its main features. 
We simulated data from a deep generative model that takes into account common features of scRNA-seq data (count distribution, library size). We sampled cells from synthetic interventions on latent variables, targeting only a sparse subset of them. 
With these simulations, we have access to (i) ground truth latent variables, required to assess disentanglement (ii) ground truth identity of intervention targets in latent space for each perturbation, required to assess target identification and (iii) simulation of out-of-domain samples. In this last case, these simulations provide a way to report performance under more complex settings (different sparsity rate or effect size between train and test data).

\paragraph{Baselines}
Because practitioners rely on disentanglement methods with the hope for recovering more insightful representations of single cell data, we include as baselines the standard vanilla VAE~\citep{kingma2013auto} and its popular derivative $\beta$-VAE~\citep{higgins2016beta}, both adopted by the biology community~\citep{lopez2018deep,eraslan2022single}. We also include the iVAE \citep{khemakhem2020variational} as well as the sVAE \citep{lachapelle2022disentanglement}, both explicitly aiming at latent units recovery. All baselines were adjusted to account for single-cell readouts within the unified framework of \verb |scvi-tools| (details appear in Appendix~\ref{app:grids}).

\paragraph{Evaluation metrics}
Based on the ground truth provided by our simulation framework, we evaluate the performance of all methods in terms of disentanglement, causal structure learning, and transferability. For assessing disentanglement, we report the Mean Correlation Coefficient (MCC), an established metric for permutation equivalence that measures the average Pearson (or Spearman) correlation coefficients between pairs of ground truth and estimated latent variable, for the best possible permutation. A high MCC means that we successfully identified the true parameters and recovered the true sources. We also report $R^2$, a metric for assessing identifiability up to a linear transformation.
To evaluate the learned causal structure, we report the precision, recall and F1 score of the learned adjacency matrix of $\hat{G}^a$ compared to the ground truth $G^a$, taking into account the permutation equivalence of $z$. 
Finally, to assess transferability, we report the negative log-likelihood of data points from holdout perturbations (Interventional NLL). For such a perturbation $a^*$, we fine-tune the model by learning parameters for the prior $p(z_n \mid a_n^*)$, while keeping the rest of the generative model $p(x_n \mid z_n)$ fixed~\citep{gentzel2019case}. In order to keep this evaluation procedure as simple as possible, we do not regularize for sparsity with sVAE and sVAE+ during the fine-tuning step. Finally, the likelihood is approximated using the importance weighted ELBO with $5,000$ particles~\citep{iwae}.

\begin{table}[t]
\centering
\caption{Mean and standard deviation per metric on simulations for $d = 15$. Best in bold. }
~\newline
\resizebox{\textwidth}{!}{%
\begin{tabular}{@{}lccc|ccc|c@{}}
\toprule 
\multicolumn{4}{c|}{ \textbf{Disentanglement}}    & \multicolumn{3}{c|}{\textbf{Causal discovery}}    & \multicolumn{1}{c}{\textbf{OOD}}   \\
  & \textbf{Pearson MCC ($\uparrow$)}  & \textbf{Spearman MCC ($\uparrow$)} & $\mathbf{R^2}$ ($\uparrow$)   & \textbf{Precision}  ($\uparrow$)& \textbf{Recall} ($\uparrow$) & \textbf{F1} ($\uparrow$) & \textbf{Inter. NLL} ($\downarrow$)\\ \midrule
\textbf{VAE}  & 0.46 $\pm$ 0.02 & $0.40 \pm 0.02$ & $0.82 \pm 0.00$ & - & - & -    &  327.10 $\pm$ 2.81 \\
\textbf{$\beta$-VAE}  & $0.48 \pm 0.02$ & $0.42 \pm 0.02$ & $0.83 \pm 0.00$ & - & - & - & 392.68 $\pm$ 4.58 \\
\textbf{iVAE} & 0.47 $\pm$ 0.01 & $0.39 \pm 0.00$ & $0.85 \pm 0.00$ & $0.16 \pm 0.02$ & \textbf{0.18} $\pm$ \textbf{0.03} & $0.17 \pm 0.08$ & $323.26 \pm 2.88 $\\
\textbf{sVAE} & $0.72 \pm 0.14$ & $0.63 \pm 0.13$ & $0.84 \pm 0.01$ & $0.43 \pm 0.19$ & $0.39 \pm 0.37$   & $0.31 \pm 0.13$  & $318.26 \pm 3.86$\\
\textbf{sVAE+} & \textbf{0.88} $\pm$ \textbf{0.04} & \textbf{0.79} $\pm$ \textbf{0.04} & \textbf{0.86} $\pm$ \textbf{0.01} & \textbf{0.54} $\pm$ \textbf{0.10} & 0.47 $\pm$ 0.09   & \textbf{0.51} $\pm$ \textbf{0.09} & \textbf{315.43} $\pm$ \textbf{2.46}\\ \bottomrule
\end{tabular}
}
%
\label{tab:synthetic_results}
\end{table}

\subsection{Experiments with synthetic data}

As part of the empirical evaluation, we explored the performance of all methods across several values of the latent space dimension, $d \in \{5, 10, 15, 20\}$. In all cases, we simulated data subject to $K = 100$ interventions, with 500 cells sampled per intervention. Measurements from 20 interventions were held-out from the training set, and used for transfer learning evaluation. All configurations were ran for 5 different random initialization. We selected the optimal hyper-parameters of each method (number of epochs, $\beta$, and sparsity penalty $\alpha$) using the Unsupervised Disentanglement Ranking (UDR) framework~\citep{duan2019unsupervised} (hyper-parameter grids appear in Appendix~\ref{app:grids}). The number of latent variables of the generative model was fixed to the one used in the simulation.

We present the results for all methods across different metrics for $d=15$ in Table~\ref{tab:synthetic_results}. An extended, systematic overview for different setups can be found in Appendix~\ref{app:results}. Our takeaways from the synthetic experiments are as follows:
\begin{itemize}
 \setlength{\itemsep}{0pt}
    \item \textbf{Disentanglement} All methods achieve high $R^2$ score, which corresponds to a satisfactory recovery of the latent units, \emph{up to a linear transformation}. However, the iVAE, sVAE and sVAE+ compare favorably to the standard VAE and $\beta$-VAE at the task of latent units recovery \emph{up to a permutation}, with sVAE+ being the strongest performer. 
    \item \textbf{Causal structure recovery} The correct structure of the causal graph can be best recovered with sVAE+, compared to sVAE and iVAE.
    \item \textbf{Out-of-distribution generalisation} sVAE and sVAE+ learning latent representations that better model data from held-out interventions, as evaluated by the Interventional NLL. 
\end{itemize}
To summarize, the proposed sandbox lets us better understand how different scenarios match with each baselines, which can be leveraged depending on the application a practitioner has in mind (whether the end goal is disentanglement, causal recovery or out-of-distribution generalization). In addition, we also notice sVAE+, that includes Bayesian treatment of the mechanism sparsity model, outperforms all VAE variants across all three objectives, without the need for tuning the sparsity hyper prior~(Appendix~\ref{sec:abl}).

\section{Empirical Evaluation on Real Datasets}
\label{sec:real_data}

We apply our benchmark models to real-world data from two recent large-scale Perturb-seq experiments~\citep{norman2019exploring,replogle2022mapping}. Further details about the processing of these data sets appear in~Appendix~\ref{app:real_world_processing}.

\subsection{In-domain Transferability and Interpretability on a Genetic Screen}

In the data set from~\cite{norman2019exploring}, $105,528$ cells from an erythrocytic leukemia  cell line (K562) were profiled after interventions targeting one or two of $112$ genes, including cell cycle regulators, transcription factors, kinases, phosphatases, and genes of unknown function. After quality control and data filtering, we retained $96,221$ cells of undergoing $212$ different genetic interventions and $8,907$ unperturbed (control) cells. Due to experimental limitations~\citep{dixit2016perturb}, we observe signal only for a subset of several thousand genes (here, $d=3,000$). The goal of the experiment was to understand the mechanisms of genetic interactions and recover gene regulatory logics. In order to simulate a transfer learning scenario, we selected the top-$30$ interventions with the most significant effect on gene expression, as assessed by the maximum mean discrepancy~\citep{mmd} estimated with a linear kernel on a PCA with dimension 50, and held out the corresponding cells as a test set. Because we did not hold out interventions according to biological knowledge, but simply based on effect size across all interventions, we qualify this benchmark of ``in-domain''. 

\begin{figure}[t]
  \centering
  \includegraphics[width=\textwidth]{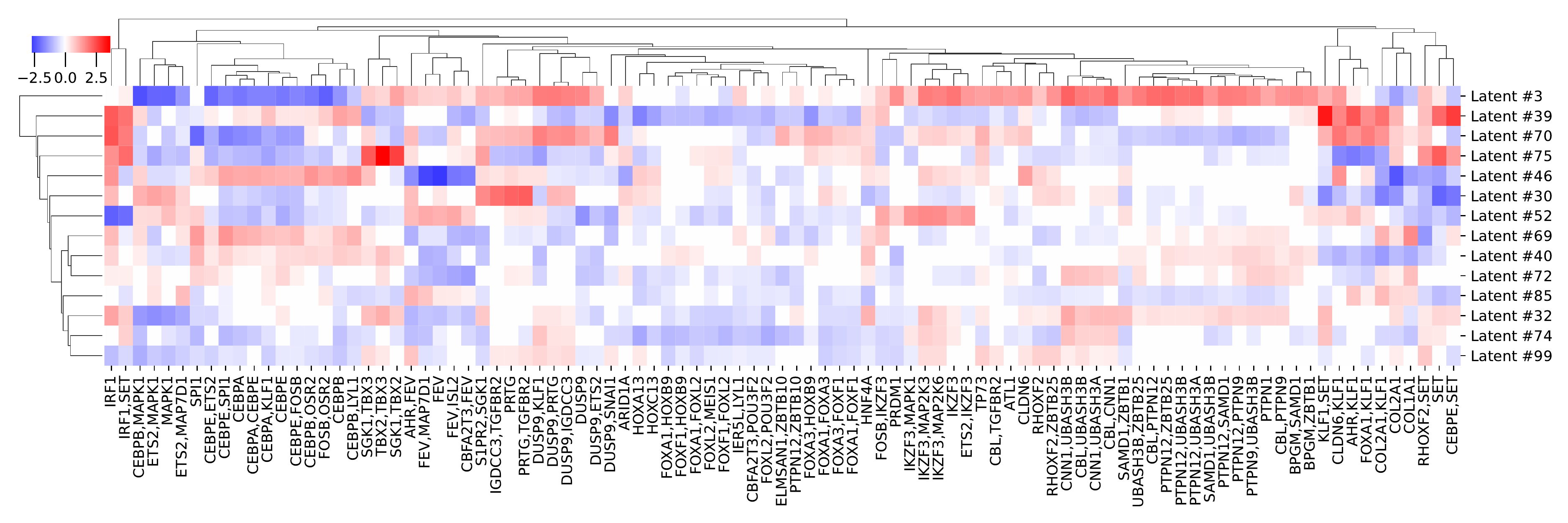}
  \caption{Perturbation effects on latent components (subset of perturbations, and components).}
  \label{fig:embeddings}
\end{figure}

We applied each studied method to this data set. Without ground truth, we use data points from held-out interventions to evaluate the models after transfer learning.
For all models, we report the negative ELBO evaluated on a validation data set (val. nELBO; including additional cells with the same perturbations as in the training data), as well the Interventional NLL (I-NLL) in Table~\ref{tab:norman}. Mean and standard deviations of the metrics are reported across five random initializations of the neural network weights. All of iVAE, sVAE and sVAE+ improve data fit compared to the VAE and $\beta$-VAE, as measured by the validation negative ELBO. However, while the iVAE provides the best fit to the validation set, with a thin margin, it fits the test set poorly compared to sVAE and sVAE+. This suggests that sVAE and sVAE+ have stronger transfer capabilities compared to other methods, and learn a more causal representation of the data.

\begin{table}[ht]
\small
\caption{\label{tab:norman}Results on the data set from~\citet{norman2019exploring}.}
~\\
\centering
\begin{tabular}{@{}lcc@{}}
\toprule
 & \textbf{val. nELBO}
 & \textbf{I-NLL}\\ \midrule
\textbf{VAE} & $717.00 \pm 0.06$ & $776.26 \pm 0.12$\\
\textbf{$\beta$-VAE} & $718.75 \pm 0.11$ & $777.18 \pm 0.12$\\
\textbf{iVAE} & $\mathbf{715.69 \pm 0.07}$ & $777.87 \pm 0.11$\\
\textbf{sVAE} & $\mathbf{715.66 \pm 0.10}$ & $776.02 \pm 0.11$\\
\textbf{sVAE+} & $716.30 \pm 0.08$ & $\mathbf{775.33 \pm 0.13}$\\
\bottomrule
\end{tabular}
\end{table}

We also performed a preliminary biological interpretation of the sparse VAE model. We first visualized the effect of perturbations in latent space in the form of a weighted adjacency matrix $W_{ij}$ for $G^a$, where the weight encodes the shift in the mean of the corresponding latent component $z_i$ for perturbation $j$ (Figure~\ref{fig:embeddings}). For visual convenience we focused on a subset of perturbations and latent components to retain the most informative data. Briefly, many perturbations have similar effects on latent variables, as it has been observed previously at the level of individual genes and programs~\citep{dixit2016perturb}. This can be interpreted as the perturbed genes being a part of the same pathway. Indeed, perturbations involving the same gene (in different combinations) grouped together by their shared effect on latent factors, as did those involving different genes from related pathways (e.g. FOXO and homeobox transcriptions factors affecting cell differentiation), whereas perturbations in genes from different pathways had different effects (e.g. IRF1 vs. CEBP family transcription factors). Additional biological interpretation of the model appears in Appendix~\ref{app:biology}.

\subsection{Out-of-Domain Transferability on a Genome-wide Genetic Screen}

\begin{figure}
    \centering
    \includegraphics[width=\textwidth]{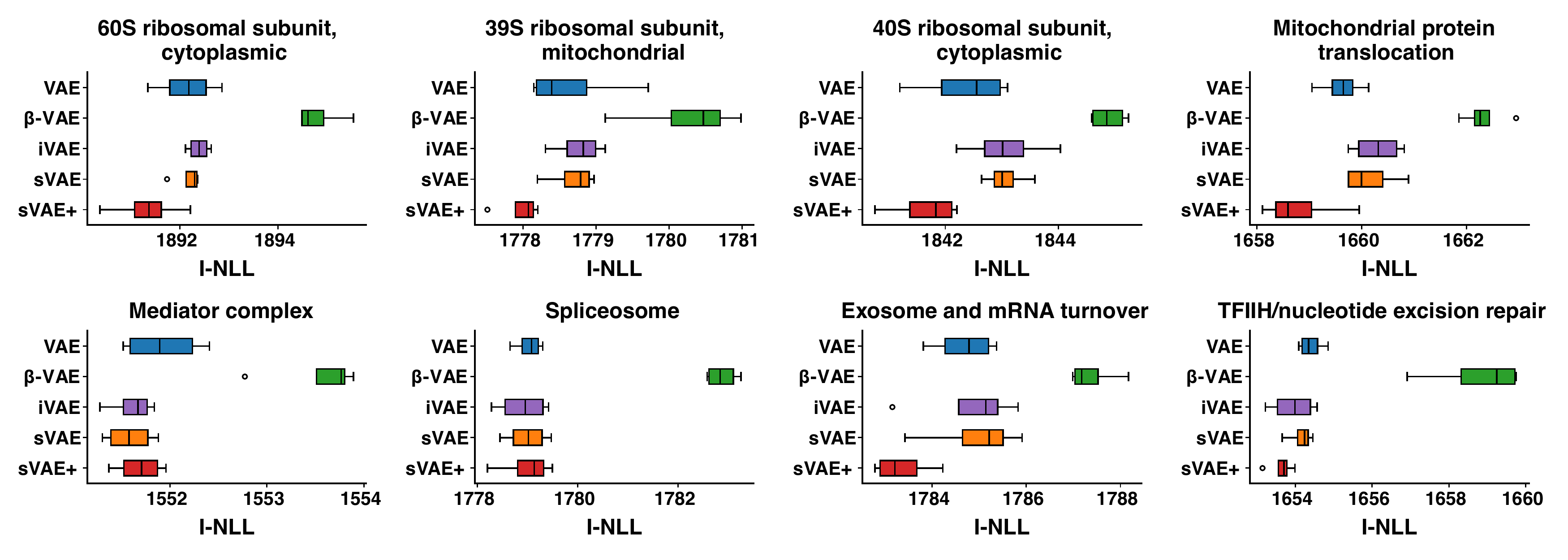}
    \caption{\label{tab:replogle}Results on the data set from~\citet{replogle2022mapping}. Each box plot reports the interventional likelihood (x-axis) on held-out interventions for a method (y-axis) trained on the full data set with one pathway hold-out.}
    \label{fig:my_label}
\end{figure}

In order to assess more systematically the transferability of the generative model learned with our benchmark methods, across distinct biological pathways, we now consider the larger data set from~\cite{replogle2022mapping}. The original data has around two million cells, after interventions targeting one of around ten thousand genes. We focused on a subset of interventions with large effect on gene expression ($K=683$), resulting in $N=116,641$ cells profiled. As in the previous data set, we observe signal only for around a thousand genes (here, $d=1,187$). Those interventions have been carefully annotated by experts into $63$ clusters, of which $8$ contained at least $20$ distinct interventions and were matched to a known biological pathway. We applied our transfer learning benchmark, treating each pathway as a set of held-out perturbations (data splits are detailed in Appendix~\ref{app:real_world_processing}), and report the interventional NLL in Figure~\ref{tab:replogle}, across five random initializations of the neural network weights. In most scenarios, sVAE+ outperforms all methods with this metric. Again, this suggests that sVAE+ learns more desirable representation of cells.


\section{Related Work}

\paragraph{Disentanglement in Variational Auto-Encoders} In order to quantify, and potentially improve upon the poorly interpretable latent variables inferred by VAEs,~\citet{higgins2016beta} created (i) a dataset with ground truth factors, (ii) metrics to quantify the mutual information between the inferred and the true factors and (iii) a modification of the VAE that outperforms existing methods with respect to these metrics. The proposed modification, the $\beta$-VAE, consists in scaling the Kullback-Leibler divergence term in the evidence lower bound with a scalar $\beta > 1$. Notably, this line of work does not necessarily associate disentanglement with the exact recovery of latent variables (unlike this paper), but is more largely concerned by the conservation of coherent axes of variation in the data. Several research groups proposed novel algorithms, such as the $\beta$-TCVAE~\citep{chen2018isolating+} or the Factor-VAE~\citep{kim2018disentangling}, while others improved benchmarking and metrics~\citep{eastwood2018a,duan2019unsupervised}.

\paragraph{Identifiable Models from Unsupervised Data} Although deep generative models are in general unidentifiable, some recent work proposed different sets of assumptions for which recovery of latent variables is possible. For example,~\citet{moran2022identifiable} proposed a VAE model for which each feature may be generated from only a sparse set of latent factors. Such model is proven to be identifiable under an anchor assumption, that stipulates that for each latent factor, there exists at least two features depending only on that factor. This hypothesis is also relevant to the field of single-cell genomics, for which a few genes may be reasonably expected to be part of only one biological process. More recent work~\citep{kivva2022identifiability} provides an identifiability result for unsupervised deep generative models under a mixture model prior.  


\paragraph{Supervised Generative Models} Both of the iVAE and the sVAE are particular instances of conditional deep generative models~\citep{sohn2015learning}. These models place the auxiliary variable $a$ ``upstream'' of the latent factors $z$ in the corresponding graphical model, in the form of a conditional prior $p(z \mid a)$. A different modeling choice would consist in placing the auxiliary variable ``downstream'' of $z$, an idea commonly exploited in supervised topic models~\citep{mcauliffe2007supervised}. Such modeling hypotheses have limited causal semantic, but may still be useful for recovering biological processes that are helpful for analyzing perturbation data, because the supervision may effectively guide inference towards topics that are more reflective of those perturbations.

\paragraph{Causal Structure Learning} Although here we consider the setting of learning from interventional data with targets being latent variables (and therefore unknown), a related line of work is concerned with learning causal relationships at the level of features, possibly under interventions~\citep{wang2017permutation}. These methods are in principle applicable to large-scale genetic screens~\citep{lopez2022large}, because by design we know which genes are targeted by each intervention. However, current algorithms for causal structure learning are very limited in their ability to handle internal cellular states (confounding factors), a task for which latent variable models are more suitable. Therefore, this paper presents an alternative modeling choice for this type of interventional data.

\section{Discussion}

We propose to explicitly model perturbations in single-cell genomics as interventions on a latent space, with a causal semantic. This naturally leads to the application of the sVAE~\citep{lachapelle2022disentanglement} and iVAE~\citep{khemakhem2020variational} framework, as well as our proposed sVAE+ model, to disentangle the latent space of single cell data by leveraging additional knowledge of perturbations. We provide a benchmarking framework for assessing the performance of the learned representations in terms of level of disentanglement, causal target identification, as well as transfer learning. In simulated data, both approaches outperform the $\beta$-VAE and the vanilla VAE, with a strong advantage for sVAE+, explicitly assuming sparsity in mechanism shifts for each perturbation. We also applied all methods to two real data sets. Our analysis suggests that sparsity may help in transfer learning and interpretation of latent variables. Importantly, we see in Figure~\ref{fig:embeddings} that multiple latent variables are affected by each intervention, which suggests more informative constraints to the model could be added to further improve its interpretability \citep{lotfollahi2022biologically}.

The hypotheses from the sVAE+ model in its current state, however, may present a few limitations for biological applications. Importantly, although it may be reasonable to expect that genetic (and often chemical) interventions directly trigger a sparse subset of a cell's circuitry (e.g., blocking a single pathway~\citep{dixit2016perturb}), there are important molecular feedback mechanisms that can induce indirect downstream effects in other pathways, especially as increasing time passes from the initial perturbation~\citep{freimer2022systematic}. Because many experiments measure gene expression from hours to days after intervention, sparsity may be a limiting assumption without resolving interactions between pathways, discussed at length in biological application of traditional causal discovery learning methods~\citep{segal2005learning,friedman2000using,pe2001inferring}. This issue could be resolved in the future with the potential availability of time-resolved measurements from single-cell perturbation experiments. We note that the treatment of those time-resolved measurements is included in the theory of~\cite{lachapelle2022disentanglement} (Theorem 5), although in this paper we focused on the more specific theorem with action-sparsity described in Theorem 22. In Appendix~\ref{app:assumption}, we discuss more technical assumptions of~\citet{lachapelle2022disentanglement}, such as covering of latent variable by the interventions, and sufficiently variability.  Briefly, we found that the benefit of models that account for action-sparsity is reduced when perturbations have small effects, or target a small subset of all the latent variables. 

To conclude, we present a first attempt at leveraging the sparse mechanism shift assumption for the purpose of learning causal representations as well as interpretable models of single-cell perturbation data. 
Looking forward, we anticipate that this line of work may unlock new perspectives to reason about perturbations as interventions on the genetic and molecular circuits that govern a cell's identity~\citep{wagner2016revealing}. This perspective is especially important for enhancing our understanding of biological processes leading to disease states, as well as proposing candidate targets and drugs.




\section*{Code Availability Statement}
We implement our new model and benchmarks using the \verb|scvi-tools| library, and release it as open-source software at \url{https://github.com/Genentech/sVAE}.

\acks{We thank S\'ebastien Lachapelle and Chandler Squires for insightful conversations about causal latent variable models. We acknowledge Kelvin Chen, Taka Kudo, Jan-Christian Huetter, Zia Khan, Basak Eraslan, and Anna Klimovskaia Susmelj for discussions about modeling single-cell perturbation data sets. Last but not least, we acknowledge the reviewers from the NeurIPS 2022 Workshop on Causal Machine Learning for Real-World Impact who provided very insightful feedback about an earlier version of this manuscript.

Disclosures: Romain Lopez, Natasa Tagasovska, Stephen Ra, and Kyunghyun Cho are employees of Genentech. Stephen Ra and Kyunghyun Cho have equity in Roche. Jonathan Pritchard acknowledges support from grant R01HG008140 from the National Human Genome Research Institute. Aviv Regev is a co-founder and equity holder of Celsius Therapeutics and an equity holder in Immunitas. She was an SAB member of ThermoFisher Scientific, Syros Pharmaceuticals, Neogene Therapeutics, and Asimov until July 31st, 2020; she has been an employee of Genentech since August 1st, 2020, and has equity in Roche.}

\bibliography{clear2023}

\newpage
\appendix
\part*{Appendices}

The appendices are organized as follows. We first provide the mathematical derivations of the evidence lower bound of sVAE+ (Appendix~\ref{app:elbo}), along with implementation details~(Appendix~\ref{app:implementation}). Then, in Appendix~\ref{app:simulation}, we provide details about our simulation framework. In Appendix~\ref{app:grids}, we explain the baselines used in this study, and the hyperparameter grid used for reporting the experimental results. In Appendix~\ref{app:results}, we provide supplementary experimental results. In Appendix~\ref{app:real_world_processing}, we report the preprocessing steps for the real-world data sets. 
In Appendix~\ref{app:assumption}, we discuss the technical assumptions from the work of~\citet{lachapelle2022disentanglement}, and the implication it has in this study. 

\section{Derivation of the Evidence Lower Bound for sVAE+}
\label{app:elbo}

We now derive the ELBO that is used as the objective function for inference within the sVAE+ framework. Let us note $\bm{X} = [x_n]_{n=1}^N$ and $\bm{A} = [a_n]_{n=1}^N$. We also adopt vector notation for the latent variables $\bm{\gamma} = [\gamma^a_i]_{a \in [K], i \in [p]}$, $\bm{\pi} = [\pi^a_i]_{a \in [K], i \in [p]}$ and $\bm{Z} = [z_n]_{n=1}^N$. We remind the reader that the likelihood of the data may be written as an intractable integral over the (latent) random variables:
\begin{align}
    \log p(\bm{X} \mid \bm{A}) &= \log \prod_{n=1}^Np(x_n \mid a_n) \\
    &= \log \iiint\prod_{n=1}^N p(x_n, z_n, \bm{\gamma}, \bm{\pi} \mid a_n)d\bm{Z}d\bm{\gamma}d\bm{\pi}\\
    &= \log \iiint \prod_{n=1}^N p(x_n \mid z_n) p(z_n \mid \gamma_{a_n}) \prod_{a \in [K]}\prod_{i \in [p]}p(\gamma^a_i \mid \pi^a_i)p(\pi^a_i)d\bm{Z}d\bm{\gamma}d\bm{\pi}. \label{eq:evidence}
\end{align}
Now, we remind the reader of the variational distribution
\begin{align}
    \bar{q} = \prod_{n \in [N]}q(z_n \mid x_n, a_n) \prod_{a \in [K], i \in [p]}q(\gamma^a_{i})q(\pi^a_{i}).
\end{align}
To derive the evidence lower bound, we start by weighting the integrand by the variational distribution as follows:
\begin{align}
    \log p(\bm{X} \mid \bm{A}) &= \log \mathbb{E}_{\bar{q}} \left[\prod_{n=1}^N \frac{p(x_n \mid z_n) p(z_n \mid \gamma_{a_n})}{q(z_n \mid x_n, a_n)} \prod_{a \in [K]}\prod_{i \in [p]}\frac{p(\gamma^a_i \mid \pi^a_i)p(\pi^a_i)}{q(\gamma^a_{i})q(\pi^a_{i})}\right],
\end{align}
and then use the concavity of the natural logarithm to apply Jensen inequality:
\begin{align}
    \log p(\bm{X} \mid \bm{A}) &\geq \mathbb{E}_{\bar{q}} \log \left[\prod_{n=1}^N \frac{p(x_n \mid z_n) p(z_n \mid \gamma_{a_n})}{q(z_n \mid x_n, a_n)} \prod_{a \in [K]}\prod_{i \in [p]}\frac{p(\gamma^a_i \mid \pi^a_i)p(\pi^a_i)}{q(\gamma^a_{i})q(\pi^a_{i})}\right].
\end{align}
From this follows the celebrated evidence lower bound (ELBO):
\begin{align}
       \log p(\bm{X} \mid \bm{A}) &\geq \mathbb{E}_{\bar{q}}\left[\sum_{n=1}^N \log \frac{p(x_n, z_n \mid \gamma_{a_n})}{q(z_n \mid x_n, a_n)} + \sum_{a \in [K], i \in [p]}\log \frac{p(\gamma^a_{i} \mid \pi^a_{i})p(\pi^a_{i})}{q(\gamma^a_{i})q(\pi^a_{i})}\right].
\end{align}
Exploiting the analytical expressions for (i) the Kullback-Leibler divergence between two multivariate Gaussian distributions, (ii) the Kullback-Leibler divergence between two Bernoulli distribution and (iii) the simplification of the expression resulting from $q(\bm{\pi}) = \delta_{\bm{\psi}}$ being a Dirac distribution, we get the final objective function:
\begin{equation}
\begin{aligned}
\label{eq:elbo}
       \log p(\bm{X} \mid \bm{A}) \geq& \mathbb{E}_{\bar{q}}\left[\sum_{n=1}^N \log p(x_n \mid z_n) - \infdiv{q(z_n \mid x_n, a_n)}{p(z_n \mid \gamma_{a_n})} \right] \\&- \sum_{a \in [K], i \in [p]} \infdiv{q(\gamma^a_{i})}{\textrm{Bernoulli}(\psi^a_i)} - \log \textrm{Beta}(\psi^a_{i}; 1, K).
\end{aligned}
\end{equation}
The last technical challenge remains in applying the reparameterization trick to $q(\bm{\gamma})$, for which we use the Gumbel-sigmoid distribution~\citep{maddison2016concrete}, as a continuous relaxation of the Bernoulli distribution. As written, the objective function may now be implemented in PyTorch ~\citep{NEURIPS2019_9015}, and \verb|scvi-tools|~\citep{gayoso2022python}. 

\section{Practical Considerations for the Implementation of sVAE+}
\label{app:implementation}

We have encountered a practical difficulty with the implementation of the evidence lower bound described in Appendix~\ref{app:elbo}. Indeed, the obtained sparsity rate in the posterior distribution $q(\gamma)$ was particularly variable with respect to the number of samples $N$, the parameter $K$ for the prior distribution $p(\pi)$ and also the number of epochs during training. This behavior potentially reveals model mis-specification, and / or sub-optimal choices for the parameterization of the variational distribution~\citep{grunwald2017inconsistency}. 

To systematically investigate this behavior, we explored re-weighting the objective function using a pseudo sample size parameter $N_\text{pseudo}$ as follows:
\begin{equation}
\begin{aligned}
\label{eq:re-elbo}
       \text{ELBO}_\text{pseudo} &= \mathbb{E}_{\bar{q}}\left[\frac{N_\text{pseudo}}{N}\sum_{n=1}^N \log p(x_n \mid z_n) - \infdiv{q(z_n \mid x_n, a_n)}{p(z_n \mid \gamma_{a_n})} \right] \\&- \sum_{a \in [K], i \in [p]} \infdiv{q(\gamma^a_{i})}{\textrm{Bernoulli}(\psi^a_i)} - \log \textrm{Beta}(\psi^a_{i}; 1, K).
\end{aligned}
\end{equation}
Using the re-weighted ELBO in~\eqref{eq:re-elbo} as the objective function, with $N_\text{pseudo} = 200$, we were able to outperform all baselines and obtain state-of-the-art results on the simulation scheme presented in Appendix~\ref{app:simulation} (not reported in this manuscript). 

However, because the optimal value of $N_\text{pseudo}$ may change depending on the data set, and more specifically on the values of $p$, $K$ and $N$, we sought to find a simpler implementation that would perform well throughout the paper with a minimum number of parameters to tune. The practical solution\footnote{This solution is implemented in the code, and the original lower bound appears as commented.} we retained was to fix the value of $\psi_i^a$ to $\mathbb{E}[q(\gamma_i^a)]$ in \eqref{eq:elbo}, instead of optimizing it as a free parameter. We have found this simplification of the inference procedure to outperform all baselines, and therefore was applied throughout the paper.

This simplification results in an inference procedure that provides no treatment of the uncertainty for the sparsity patterns in the graph $\hat{G}^a$. This is an important point, because such measure of uncertainty is insightful to assess the statistical significance of the edges in the causal graph (following the principles of Bayesian decision theory, with examples in~\cite{Clivio794875} and \cite{lopez2020decision}). We therefore expect that more advanced treatment of the sparsity, such as improvements over the Beta-Bernoulli prior, or in the inference procedure, will provide substantial improvement as well as usability to sVAE+. We treat such practical advances as future work.

\section{Simulation Details}
\label{app:simulation}
We simulate single-cell gene expression profiles from perturbation experiments as follows. We assume we have measurements from $N$ cells. Each of the cells, for example cell $n \in [N]$, has been exposed to a perturbation/intervention $a_n \in [K]$. We use $K = 100$ interventions in our simulations, and sample $N = 100,000$ cells in total. The first $80$ interventions form the training set, and the last $20$ form the test set. Each intervention $a \in [K]$ is represented by a sparse perturbation embedding $\mu_{a} \in \mathbb{R}^p$, where $p=15$ is the dimension of the embedding. 

For each intervention, we treat the number of affected latent dimensions $t_a = \{1, 2, 3\}$ uniformly at random. The indices of affected dimensions are also drawn without replacement from $[p]$, encoded into a binary vector $\beta_{a, .} \in \{0, 1\}^p$. Finally, each component $\mu_{a, i}$ of $\mu_a = \left( \mu_{a, 1}, \ldots, \mu_{a, p}\right)$ for $i\in [p]$ is generated as:
\begin{align}
 \eta_{a, i} &\sim \frac{1}{2}\textrm{Normal}(-e, 0.5) + \frac{1}{2}\textrm{Normal}(e, 0.5)\\
 \mu_{a, i} &\sim (1- \beta_{a,i}) \delta_0 + \beta_{a, i} \eta_{a, i},
\end{align}
where $e$ is a scalar quantifying the strength of the intervention, and $\delta_0$ designates the Dirac delta distribution with mass at $0$. For cell $n$, exposed to intervention $a_n$, latent variable $z_n$ is generated as:
\begin{align}
 z_n \sim \textrm{Normal}(\mu_{a}, I),
\end{align}
where each individual component of $z_n$ encodes the activity of a pathway, shifted by the intervention. Measurements $x_{ng}$ from a single cell $n$ and a gene $g$ are generated from a count distribution:
\begin{align}
 x_{ng} \sim \textrm{Poisson}\left(l_n f^g(z_n) \right),
\end{align}
\label{eq:obs}
where $l_n$ is the library size fixed to $10^5$, and the mixing function $f$ is a neural network with three hidden layers of 40 units, Leaky-ReLU activations with a negative slope of 0.2, and a softmax non-linearity on the last layer to convert the outputs to counts~\citep{lopez2018deep}. The weight matrices of $f$ are sampled according to an isotropic Gaussian distribution, with orthogonal columns, to make sure $f$ is injective~\citep{lachapelle2022disentanglement}. Although we used a Poisson distribution for simplicity, future work will investigate the use of more realistic count distributions such as Beta-Poisson, or negative binomial, that may be important to model over-dispersion of scRNA-seq data~\citep{zhang2019simulating}.

\section{Details for Empirical Evaluation of the Benchmark Methods}
\label{app:grids}
In this section, we provide the necessary details for reproducing the experiments in the paper.

\subsection{Baseline and Metric Details}

\paragraph{Shared components of the generative model} All models rely on the generative model described in Section~\ref{sec:ssm_sc}. For the VAE and $\beta$-VAE, $\bm{\gamma} = \bm{0}$ while for iVAE, $\bm{\gamma} = \bm{1}$. Similarly, the sVAE is implemented with a Laplace prior on $\bm{\pi}$ as in the original publication. The neural network architecture for $f$ in the generative model is given in \autoref{app:simulation}. 

\paragraph{Shared components of the inference} The mean and variance of the variational distribution for $z$ are each obtained as the output of a neural network with two hidden layers, 128 hidden units at each layer and ReLU non-linearity in between hidden layers, taking as input the observed data $x$. We optimize the ELBO using the Adam optimizer~\citep{kingma2014adam} with minibatches of size 128.

\paragraph{VAE and $\beta$-VAE} Since both of these models do not have the ability to use the interventions while modeling the data, in our implementation we let them use all the available data and simply ignore the covariate~$a$. 

\paragraph{iVAE} Unlike sVAE and sVAE+ which can estimate a causal graph, iVAE only outputs estimates for the parameters $\bm{\mu}$ (effect of each intervention on each latent variable).  In order to compare those three baselines, we need a heuristic to build an adjacency matrix from the estimates of $\bm{\mu}$. In this benchmark, we initially used the outlier detection method from~\citet{rousseeuw1999fast}, based on the idea that for each intervention, we seek to find the latent variables with most significant sensitivity to it. However, the method have yielded poor result (high recall), therefore we simply selected the top-2 latent units per intervention, ordered by the absolute value of their $\mu$ parameter. With such type of heuristics, we ensure to have a sparse graph, based on the most meaningful factors. 

\paragraph{Transfer Learning Details} For the transfer learning experiments, we fix the parameters of the generative model, as well as the ones of the variational network. For VAE and $\beta$-VAE, we report the importance weighted ELBO (IWELBO) for the hold-out interventions. For iVAE, sVAE, and sVAE+, we proceed to a fine-tuning step with all of the above parameters frozen, except for the parameter $\bm{\mu}$. In this fine-tuning step, we fix the variables $\bm{\gamma}$ to $\bm{1}$ to avoid the sparsity prior to constrain the model in fitting the test set. The IWELBO~\citep{iwae} with $T$ particles is calculated as:
\begin{align}
       \textrm{IWELBO}^T(\bm{X} \mid \bm{A}) &= \sum_{n=1}^N \mathbb{E}_{z_n^1, \ldots, z_n^T \sim q(z_n \mid x_n, a_n)}\left[\log \frac{1}{T}\sum_{t=1}^T \frac{p(x_n, z^t_n \mid \gamma_{a_n} = 1)}{q(z^t_n \mid x_n, a_n)} \right].
\end{align}

\subsection{Hyper parameters Grid for the Simulated Data}
For all methods, we performed an exhaustive hyper parameter grid search. We used the Unsupervised Disentanglement Ranking (UDR) framework~\citep{duan2019unsupervised} for selecting the optimal hyper parameters. The complete hyper parameter search space for each algorithm is described in Table~\ref{tab:hyperparams}.

\begin{table}[ht]
\centering
\caption{Grid search spaces for each baseline.}~\\~
\begin{tabular}{@{}ll@{}}
\toprule
\textbf{}     & \textbf{Hyperparameter space} \\ \midrule
\textbf{VAE}  & $n_{\text{epochs}}  \in   \{300, 500\}$     \\  
$\beta$-\textbf{VAE}  & \begin{tabular}[c]{@{}l@{}}$n_{\text{epochs}}  \in   \{300, 500\}$ , $\beta  \in  \{8, 10, 30\}$\end{tabular}            \\   
\textbf{iVAE} & $n_{\text{epochs}}  \in   \{300, 500\}$     \\  
\textbf{sVAE}  & \begin{tabular}[c]{@{}l@{}}$n_{\text{epochs}}  \in   \{300, 500\}$, $\alpha  \in  \{10, 20, 40, 60, 80, 100\}$\end{tabular} \\  
\textbf{sVAE+} & \begin{tabular}[c]{@{}l@{}}$n_{\text{epochs}}  \in   \{300, 500\}$ $\alpha  \in  \{10, 20, 40, 60, 80, 100\}$\end{tabular}                               \\ \bottomrule
\end{tabular}
\label{tab:hyperparams}
\end{table}

\section{Additional Results from Synthetic Experiments}
\label{app:results}



For quantitative comparison of the considered baseline methods, we leverage the availability of ground truth latent variables in simulated data sets. In the proposed sandbox in Section~\ref{sec:benchmark}, we have control over: the size of the effect for interventions - $e_a$, the sparsity of the interventions - $t_a$, as well as the dimensionality of the latent and auxiliary variables. 
The aim of this paper is to evaluate the extent to which sparse, identifiable methods provide disentangled, or ultimately ``generalizable/transferable'' representations.
To systematically assess the quality i.e. transferability of the learned representations, we propose the following scenarios:
\begin{itemize}
    \item transfer to unseen, in-domain interventions  (hold-out interventions form a data set with same level of sparsity and effect size);
    \item transfer to unseen out-of domain interventions (hold-out interventions form a data set with different level of sparsity or effect size).
\end{itemize}

\subsection{Results for different latent dimensions}
We present the disentanglement results for alternate number of latent dimensions in our simulations in Table~\ref{tab:mcc_wrt_dim}. The conclusions in the main text are robust to this choice.

\begin{table}[h]
\footnotesize
\caption{Pearson MCC scores on hold out interventions with respect to the number of latent space dimensions.} ~\\
\centering
\begin{tabular}{l|cccccccccccccccc}
\toprule
\textbf{} & \textbf{VAE} & \textbf{iVAE} & \textbf{sVAE} & \textbf{sVAE+} & \textbf{$\beta$-VAE} \\
\midrule
$p= 5$ & $0.59 \pm 0.12$ & $0.69 \pm 0.09$ & 0.71 $\pm$ 0.09 & \textbf{0.94} $\pm$ \textbf{0.03} & $0.63 \pm 0.06$ \\
$p= 10$ & $0.55 \pm 0.10$ & $0.66 \pm 0.13$ & 0.72 $\pm$ 0.13 & \textbf{0.91} $\pm$  \textbf{0.03}& $0.58 \pm 0.03$ \\
$p= 15$  & $0.56 \pm 0.09$ & $0.68 \pm 0.11$ & 0.72 $\pm$ 0.08 & \textbf{0.88} $\pm$ \textbf{0.04} & $0.58 \pm 0.08$ \\
$p= 20$  & $0.47 \pm 0.01$ & $0.59 \pm 0.03$ & 0.70 $\pm$ 0.09 & \textbf{0.76} $\pm$ \textbf{0.02} & $0.55 \pm 0.08$\\
\bottomrule
\end{tabular}
\label{tab:mcc_wrt_dim}
\end{table}

\subsection{Unseen In-domain interventions}
The results included in the main section of the paper correspond to the in-domain scenario. Namely, we generate a sample data set with sparse interventions, we train each baseline on cells affected by only 80 of those interventions, and we evaluate on the hold out, not seen 20 interventions. Here, we extend this results to different dimensionality of the latent variable. 
In Table~\ref{tab:in_domain_sp}, we include the results for in-domain interventions with same levels of sparsity in both train and test sets. That is, we train a model on a data set where a fraction $s^\prime$ of all possible edges in $G^a$ are included, and we test on a data set with same sparsity. In Table~\ref{tab:in_domain_ef}, we investigate how those results hold with different effect sizes (encoded by the absolute value of the mean of the Gaussian prior shift under intervention).
Our experiments for the in-domain interventions provide the following takeaways:
\begin{itemize}
    \item observing the interventional NLL scores, the denser the ground truth latent model is, the more difficult it is to fit it with the sparse baselines
    \item observing the MCC scores, disentanglement is easier to achieve in smaller dimensionality of the latent space and when sparse shifts are present 
\end{itemize}

\begin{table}[p]
\caption{Results for transferability for in-domain interventions at different levels of sparsity $s$ of the adjacency matrix $G^a$. The effect size and latent dimension are kept fixed: $e^a = 5$ for all interventions and $d_z=10$. Results in bold are best per metric.}
~\\
\footnotesize
\centering
\begin{tabular}{l|cc|cc|cc}
\toprule
\textbf{} & \multicolumn{2}{c|}{$s' = {0.2}$} & \multicolumn{2}{c|}{$s' = {0.5}$} & \multicolumn{2}{c}{$s' = {0.9}$} \\ 
\textbf{} & \textbf{pMCC} & \textbf{Interv. NLL}  & \textbf{pMCC} & \textbf{Interv. NLL} &  \textbf{pMCC} & \textbf{Interv. NLL}  \\ \midrule
\textbf{iVAE} &0.53  & 319.27  & \textbf{0.58} & 330.56  & \textbf{0.60}   & 333.28  \\
        \textbf{sVAE} & 0.79 & 318.69 & \textbf{0.58}  & 329.68 & 0.52  & 333.98  \\
\textbf{sVAE+} & \textbf{0.83} & \textbf{309.45}    & 0.57 &  \textbf{322.01}   & 0.59  & \textbf{322.51}   \\ 
\bottomrule
\end{tabular}
\label{tab:in_domain_sp}
\end{table}

\begin{table}[p]
\footnotesize
\centering
\caption{Results for transferability for in-domain interventions for different sizes of the shift effect $e$. The sparsity size and latent dimension are kept fixed: $s^a = 0.2$ for all interventions and $d_z=5$. Results in bold are best per metric.}~\\
\begin{tabular}{l|cc|cc|cc}
\toprule
\textbf{} & \multicolumn{2}{c|}{$e^\prime =1$} & \multicolumn{2}{c|}{$e^\prime =2$} & \multicolumn{2}{c}{$e^\prime =5$} \\ 
\textbf{} & \textbf{pMCC} & \textbf{Interv. NLL}  & \textbf{pMCC} & \textbf{Interv. NLL}  & \textbf{pMCC} & \textbf{Interv. NLL} \\  \midrule
\textbf{iVAE}  & \textbf{0.78} &341.08  &0.68  & 356.84  & 0.67 &381.62  \\ 
\textbf{sVAE} & \textbf{0.78} & 341.03  & 0.68 & 349.55  &0.66  & 375.80  \\ 
\textbf{sVAE+} & \textbf{0.78} &  \textbf{338.78}  &\textbf{0.73}  & \textbf{339.80}  & \textbf{0.92} & \textbf{341.40}   \\ \bottomrule
\end{tabular}
\label{tab:in_domain_ef}
\end{table}

\subsection{Unseen Out-of-domain interventions}
Here, we use the flexibility of our sandbox to simulate different scenarios to mimic different interventions regarding change in effect size from train to test set, or change in sparsity of the bipartite graph $G^a$.
In Table~\ref{tab:out_domain_sp}, we include the results for out-of-domain interventions with different levels of sparsity between train and test sets. That is, we train a model on a data set where $s^\prime$ of all possible edges in $G^a$ are included, and we test on a data set with reduced sparsity, $s^{\prime\prime}$, s.t. $s^{\prime\prime} \geq s^{\prime}$.
Similarly, we include results for different effect sizes in Table~\ref{tab:out_domain_ef}. We train all models on a smaller effect interventions $e^\prime$ and test on interventions with larger effects $e^{\prime \prime}$.
Our experiments for the out-of-domain interventions provide the following takeaways:
\begin{itemize}
    \item observing the interventional NLL scores, the more sparse the ground truth latent model in the transfer domain, the more difficult it is to fit it for all baselines 
    \item observing the MCC scores, disentanglement, or identifiability of the true latent variables is not possible in the out-of-domain interventions.
\end{itemize}

\begin{table}[p]
\footnotesize
\centering
\caption{Results for transferability for out-of-domain interventions at different levels of sparsity $s$ of the adjacency matrix $G^a$. The effect size and latent dimension are kept fixed: $e^a = 5$ for all interventions and $d_z=10$. Results in bold are best per metric.}
~\\
\begin{tabular}{l|cc|cc|cc}
\toprule
\textbf{} & \multicolumn{2}{c|}{$s' ={0.2} \rightarrow s'' ={0.5}$} & \multicolumn{2}{c|}{$s' = {0.5} \rightarrow s'' = {0.7}$} & \multicolumn{2}{c}{$s' = {0.7} \rightarrow s'' = {0.99}$} \\
\textbf{} & \textbf{pMCC} & \textbf{Interv. NLL}  & \textbf{pMCC} & \textbf{Interv. NLL}  & \textbf{pMCC} & \textbf{Interv. NLL}  \\ \midrule
\textbf{iVAE} &\textbf{0.44} & 667.77  & 0.38 & 721.55 & 0.44  &  762.35  \\

\textbf{sVAE} & 0.40 & 676.64 & 0.38  & \textbf{720.33}   & 0.37 & 710.88    \\
\textbf{sVAE+} & 0.41 & \textbf{656.61}  & \textbf{0.42} & 675.15 & \textbf{0.49}   & \textbf{681.25}    \\
\bottomrule
\end{tabular}
\label{tab:out_domain_sp}
\end{table}

\begin{table}[p]
\footnotesize
\centering
\caption{Results for transferability for out-of-domain interventions for different sizes of the shift effect $e$. The sparsity size and latent dimension are kept fixed: $s^a = 0.2$ for all interventions and $d_z=5$. Results in bold are best per metric.}
~\\
\begin{tabular}{l|cc|cc|cc}
\toprule
\textbf{} & \multicolumn{2}{c|}{$e^\prime=1 \rightarrow e^{\prime\prime}=3 $} & \multicolumn{2}{c|}{$e^\prime=2 \rightarrow e^{\prime\prime}=4 $} & \multicolumn{2}{c}{$e^\prime=5 \rightarrow e^{\prime\prime}=7 $} \\ 
\textbf{} & \textbf{pMCC} & \textbf{Interv. NLL}  & \textbf{pMCC} & \textbf{Interv. NLL} & \textbf{pMCC} & \textbf{Interv. NLL}  \\  \midrule
\textbf{iVAE} & \textbf{0.47} &1179.31 & \textbf{0.59}  & 1153.24 & \textbf{0.58}   &1224.26   \\

\textbf{sVAE} & \textbf{0.47} & 1173.11  & 0.44 & 1168.79  &0.43  & 1229.12  \\
\textbf{sVAE+} &\textbf{0.47}  & \textbf{1103.91}   & 0.51 &\textbf{1104.70}  & 0.46 &  \textbf{1179.39}   \\
\bottomrule
\end{tabular}
\label{tab:out_domain_ef}
\end{table}

\subsection{Additional ablation analysis}
\label{sec:abl}
\begin{figure}[ht!]
  \centering
  \includegraphics[width=0.95\textwidth]{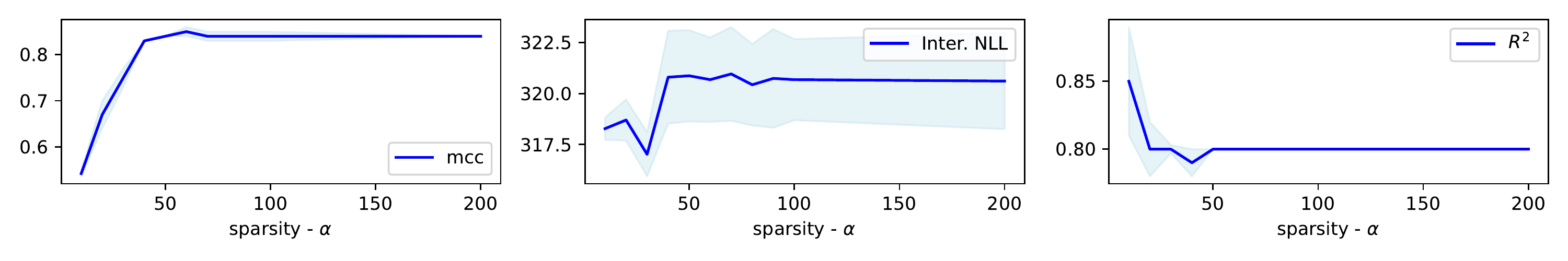}
  \caption{Pearson MCC, interventional NLL and $R^2$ score with respect to different values for the sparsity hyperparameter $\alpha$ in sVAE+.}
  \label{fig:ssvae_abl}
\end{figure}

Additionally, we explored the robustness of sVAE+ with regards to the hyper prior $\alpha$ defining the shape parameters $(1, \alpha)$ for the Beta distribution. We remind the reader that this distribution controls the level of sparsity of the causal graph $G^a$ estimated by sVAE+. 
From Figure~\ref{fig:ssvae_abl}, we notice the robustness of sVAE+ with regards to different values $\alpha$. In practice, we found that choosing $\alpha=K$ (the number of interventions) has competitive performance. 

\section{Preprocessing of real world single-cell data}
\label{app:real_world_processing}

\subsection{Norman data set}
The original data set from~\cite{norman2019exploring} is publicly available from GEO (GSE133344). For this manuscript, we downloaded the data processed according to~\cite{lotfollahi2021compositional}. Cells with guide ``NegCtrl1\_NegCtrl0\_NegCtrl1\_NegCtrl0'' were excluded. All unperturbed cells were merged as one control (observational) condition. $d=3,000$ highly variable genes were selected using \verb|scanpy|~\citep{wolf2018scanpy} for training. For this data set, all models were run with $p=50$ latent variables, and for $n_{\text{epochs}}=300$ epochs. The remainder of the hyperparameters were selected via the evidence lower bound on a validation set, that contained the same interventions as the ones of the training set, as in~\citet{brouillard2020differentiable}. 

\subsection{Replogle data set}
Pre-filtered single-cell expression data of K562 cells from~\citet{replogle2022mapping} was downloaded from \href{http://gwps.wi.mit.edu}{http://gwps.wi.mit.edu}. Selection of genes and interventions, as well as clustering of interventions into pathways was obtained from Supplementary Table 3 of the original paper. The number of interventions / cells used in the training and testing data sets are detailed for each data split in Table~\ref{tab:splits}. For this data set, all models were run with $p=50$ latent variables, and for $n_{\text{epochs}}=100$ epochs. The remainder of the hyperparameters were selected via the evidence lower bound on a validation set, that contained the same interventions as the ones of the training set, as in~\citet{brouillard2020differentiable}. 

\begin{table}[ht]
\small
\centering
\caption{\label{tab:splits}Number of interventions used across data splits used for the Replogle dataset.}
~\\
\begin{tabular}{lll}
\toprule
                                     & $K_\text{train}$ & $K_\text{test}$ \\ \midrule
\textbf{Exosome and mRNA turnover}            & 663      & 20      \\
\textbf{Spliceosome}                          & 648      & 35      \\
\textbf{Mediator complex}                     & 657      & 26      \\
\textbf{TFIIH/nucleotide excision repair }    & 660      & 23      \\
\textbf{39S ribosomal subunit, mitochondrial} & 586      & 97      \\
\textbf{60S ribosomal subunit, cytoplasmic}   & 640      & 43      \\
\textbf{40S ribosomal subunit, cytoplasmic}   & 630      & 53      \\
\textbf{mitochondrial protein translocation } & 643      & 40     \\ \bottomrule
\end{tabular}
\end{table}

\section{Biological interpretation of the sVAE+ model on the Norman dataset}
\label{app:biology}

\begin{figure}[htb]
    \begin{minipage}[t]{.5\textwidth}
        \centering
        \includegraphics[width=0.7\textwidth]{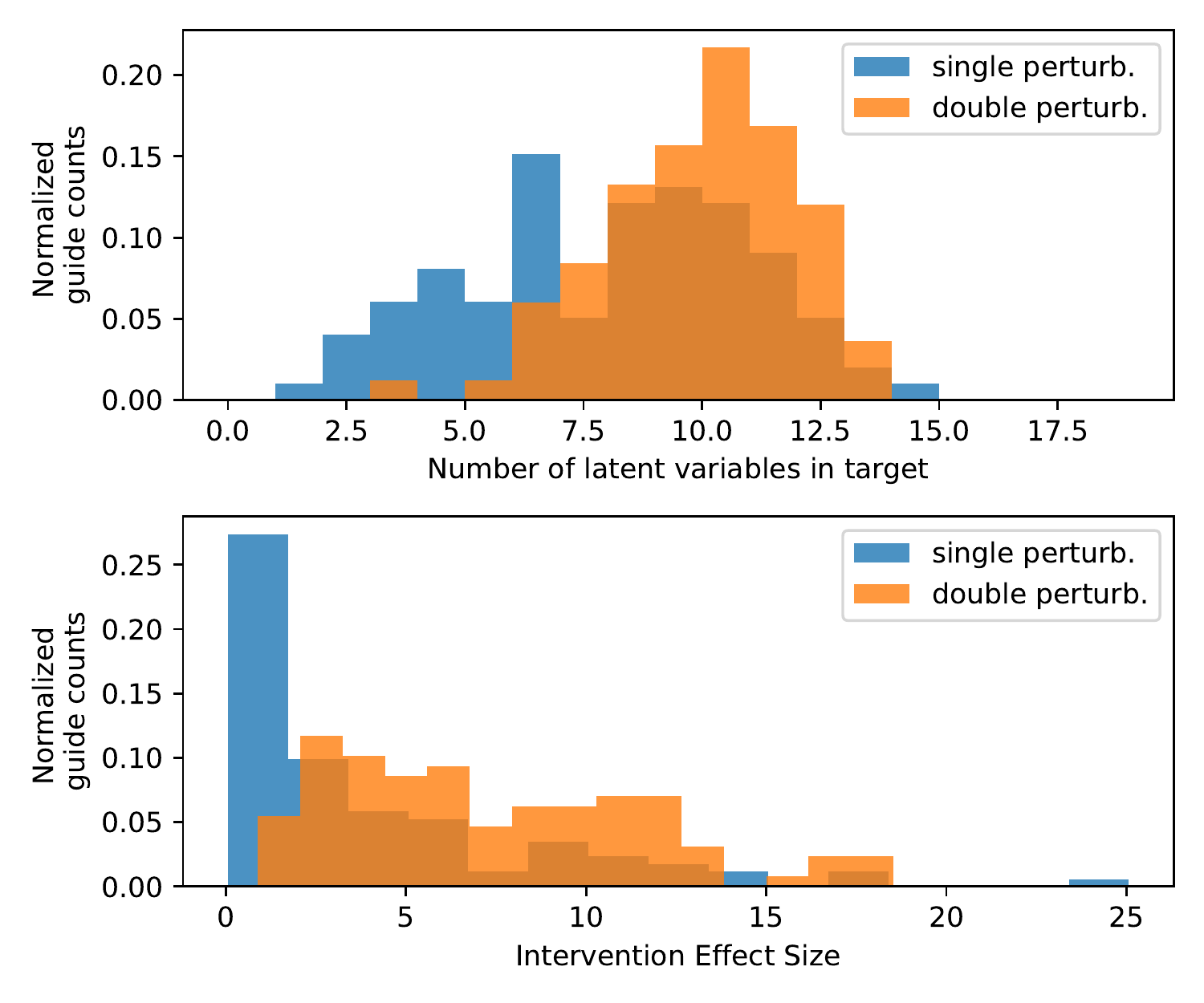}
    \end{minipage}
    \hfill
    \begin{minipage}[t]{.5\textwidth}
        \centering
          \includegraphics[width=0.7\textwidth]{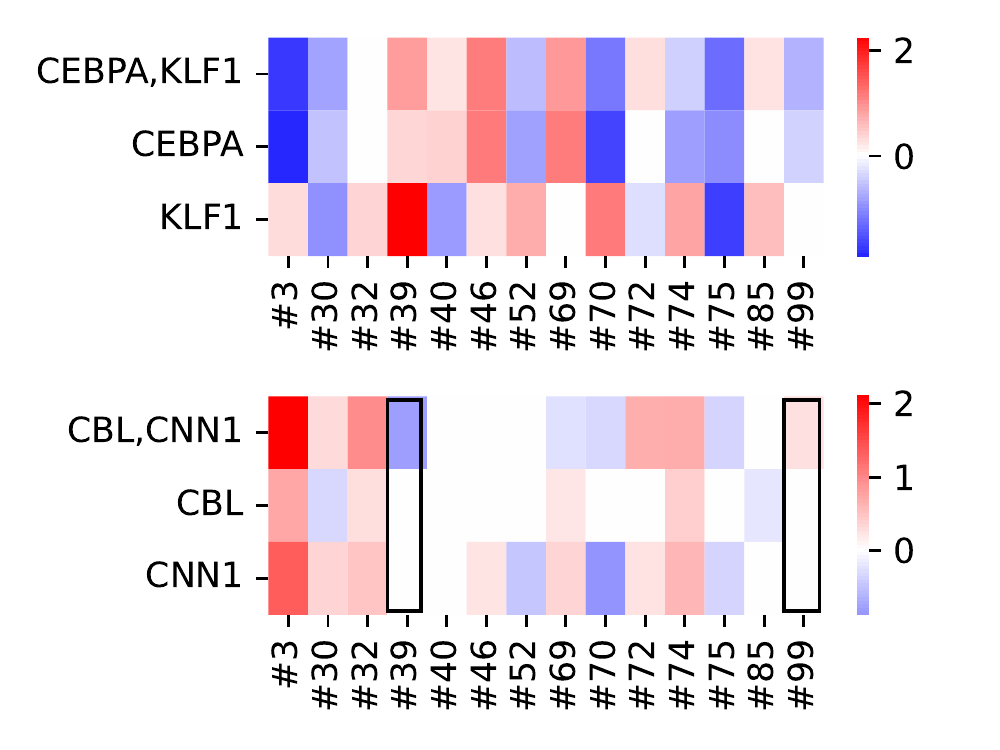}
    \end{minipage}
    \caption{(A) Targets and effect size identified by the model. (B) Genetic interactions as identified from target and effect sizes with the model.\label{fig:subset}}
\end{figure}

We investigated how statistics of the number of perturbed latent variables and/or of the effect size was changing according to whether  one or two genes were targeted in the cell (Figure~\ref{fig:subset}a). The distribution of both statistics is significantly higher for double vs. single gene perturbations, as overall expected. 

Finally, we sought to assess whether the learned latent variables are reflective of known patterns in genetic interactions. Two examples of genetic interactions are pointed out in Figure~\ref{fig:subset}b. In the first one, we may notice that the latent shift for the perturbation that involved a combination of perturbations in CEBPA and KLF1 has a pattern mostly similar to the shift of a single gene perturbation CEBPA, as previously reported~\citep{norman2019exploring}. This is an example of a dominant interaction, already visible in Figure~\ref{fig:embeddings}, in other combinations (e.g., DUSP9, ETS2). In the second example (CLB, CNN1), the sparsity pattern identifies two latent variables (number 39 and 99; black rectangle) with a shift that did not appear in any of the individual perturbations. We applied Integrated Gradients~\citep{sundararajan2017axiomatic} to each of those two components of the encoder network to obtain a list of 50 most important genes, and used EnrichR~\citep{chen2013enrichr} to obtain an associated gene signature. A positive change in latent variable 99 was associated with hemoglobin alpha binding, and hydrogen peroxyde metabolic process, both important in the context of erythocytes (the source cell line). Latent variable 39 was associated with RNA binding.

\section{Discussion on identifiability (sVAE) assumptions in single-cell data}
\label{app:assumption}

The original version of sVAE as proposed in \citet{lachapelle2022disentanglement} does not account for single-cell data. Hence we did modifications to adjust for that, as explained in Section~\ref{sec:ssm_sc}. Additionally, for the identifiability theory of sVAE  to hold, a number of assumptions are required, as stated in their Theorem 5. In what follows we discuss the main differences and modifications in the single-cell context. 

\subsection{Discrete observation space} The basic theory of~\cite{khemakhem2020variational}, and~\cite{lachapelle2022disentanglement} assumes that the data $x$ is generated as $x = f(z) + \epsilon$ where components of $\epsilon$ are mutually independent, and all independent of $z$. It also assumes that $f$ is a differentiable bijection with a differentiable inverse. Consequently, this theory may not be used to back up the claim that the decoder $f$ of the generative model detailed in Section~\ref{sec:ssm_sc} is identifiable. Indeed, we assume that $x$ follows a count distribution whose parameters depend on $z$, and this goes beyond the framework described by the aforementioned theoretical papers. However, the appendices of~\cite{khemakhem2020variational} (as well as the Section~\ref{sec:ssm_sc} of this paper) describe encouraging empirical results that algorithms motivated by theory for data with linear Gaussian observation noise also improve performance on discrete observation spaces. The characterization of the theoretical conditions that are necessary and sufficient to guarantee identifiability with a discrete observation space remains an open problem in the field.

\subsection{Sufficient variability} Both sVAE and iVAE require an ``assumption of variability'' that specify that the conditional distribution $p(z \mid a)$ must sufficiently vary with $a$. In both papers, this assumption is detailed as a geometric condition and cannot be  validated in real data. Interestingly, this assumption is also not sufficient for guarantee correct estimation of the parameters in the case of finite sample sizes. This opens the way for new theoretical developments, based of non-asymptotic bounds~\citep{wainwright2019high}. For example, we may wonder how many measurements are needed, under ``sufficient variability'', for estimating the latent variables when the interventional effect size becomes small. 

From a more practical perspective, we accounted for this in our simulations by making sure the strength of the shift (encoded by parameter $e$ in Appendix~\ref{app:simulation}) was large enough. Interestingly, as we see from the simulation, when the strength of the effect is low, methods have difficulties in disentangling the latent factors~(Table~\ref{tab:out_domain_ef}).

This has also an important implication for real-world applications. In an earlier iteration of this project, we tried to apply iVAE and sVAE to the large-scale chemical screen presented in~\citet{srivatsan2020massively}, but none of those baselines outperformed the vanilla VAE. After investigation, we attribute this to (i) a majority of chemicals having low effect on gene expression, (ii) a large fraction of the variance of the data being attributed to other signals, such as cell type variation, or cell cycles. In this case, all conditional VAEs may ignore information from the covariates and produce a similar result than the vanilla VAE. 

\subsection{Graphical criterion} 
The theoretical results of~\citet{lachapelle2022disentanglement} hold under a precise condition on the structure of the ground truth graph $G_a$. For example, it is necessary (but not sufficient) for the intervention set to at least cover all of the latent variables in the model. If this is not the case, then we may have disentanglement only in a block of latent variables~\citep{lachapelle2022partial}. This assumption is particularly relevant because experiments may only focus on subsets of possible perturbations (due to cost and labor limitations) and not all latent variables may be impacted. 

Importantly, perturbing all the latent variables is not sufficient. A complete graphical condition is exhibited the sVAE paper. However, this assumption may be difficult to validate or verify in practice. Still, according to the results of our simulation experiments (that did not take into account this condition while creating $G_a$, as well as the empirical evaluation in Appendix B.4. in \citet{lachapelle2022disentanglement}, the causal graph can be approximate recovered with competitive performance compared to baselines even in the case when the graphical criterion is violated. 

We investigated this more systematically using our simulation setting, in the same conditions than in Table~\ref{tab:synthetic_results}, but by constraining the perturbations to cover only a specific number of latent variables (varying from 1 to 15) out of the 15 latent variables. Those results, reported in Table~\ref{tab:mcc_cover}, indicate that performance lower considerably when a large fraction of latent variables are not intervened upon, as expected from the theory.

\begin{table}[ht]
\centering
\caption{\label{tab:mcc_cover}Evolution of Pearson MCC metric for sVAE+ when the interventions only cover a strict subset of latent units ($d$=15).}
\begin{tabular}{cc}
\toprule
\begin{tabular}[c]{@{}c@{}}\textbf{\# targeted} \\ \textbf{latent variables}\end{tabular} & \begin{tabular}[c]{@{}c@{}}\textbf{Pearson} \\ \textbf{MCC}\end{tabular} \\ \midrule
1                                                                                      & 0.5                                                    \\
4                                                                                      & 0.59                                                   \\
7                                                                                      & 0.71                                                   \\
12                                                                                     & 0.85                                                   \\
15                                                                                     & 0.88                                                 \\ \bottomrule 
\end{tabular}
\end{table}

\section{Noise models for scRNA-seq}

Gene expression data captured by single-cell RNA sequencing can be composed as technical noise, as well as biological signal. The technical noise is complex, but usually described by two components: (a) the number of RNA transcripts captured in a single-cell is (mostly) treated as an artifact of the assay, and must be factored out of the learned representation. (b) the data takes the form of counts, with overdispersion. The best noise distribution is (still) a topic of active research, however, there is a general consensus that the noise model presented by ~\citep{lopez2018deep}, with a scaled negative binomial distribution provides satisfactory results. For more details on the noise in scRNA measurements, we refer the reader to~\cite{grun2014validation}.

\end{document}